\documentclass[reprint,
superscriptaddress,
nofootinbib,
 amsmath,amssymb,
 aps,
 prb,
]{revtex4-2}

\usepackage{graphicx}
\usepackage{upgreek}
\usepackage{epigraph}
\usepackage{hyperref}

\newcommand{\Pe}{\mathrm{Pe}}
\newcommand{\Wi}{\mathrm{Wi}}
\newcommand{\De}{\mathrm{De}}

\newcommand{\Bi}{\mathrm{Bi}}

\newcommand{\keyterm}[2]{\footnote{\textbf{#1:} #2}\ }

\begin{document}

\preprint{APS/123-QED}

\title{Life in a tight spot: Coupled dynamics of bacteria and soil across scales}

\author{Pablo Bravo}
\thanks{These authors contributed equally and are listed alphabetically by surname; their order is interchangeable.}
\author{Tanumoy Dhar}
\thanks{These authors contributed equally and are listed alphabetically by surname; their order is interchangeable.}
\author{Eloise Masquelier}
\thanks{These authors contributed equally and are listed alphabetically by surname; their order is interchangeable.}
\author{Danielle Sclafani}
\thanks{These authors contributed equally and are listed alphabetically by surname; their order is interchangeable.}
\author{Sujit S. Datta}
\email{ssdatta@caltech.edu}
\affiliation{Division of Chemistry and Chemical Engineering, California Institute of Technology, Pasadena, California 91125, USA}

\date{\today}
\begin{abstract}
\noindent 
Soil harbors much of Earth's bacterial life. The activity of these bacteria governs plant growth, carbon and nitrogen cycling, and the response of land to a changing climate. Understanding this activity is difficult, however: soil is structurally and chemically heterogeneous and optically opaque, and its bacteria not only respond to their surroundings but continually reshape them, a two-way feedback that most idealized experiments and theories overlook. Here we review how this dynamic feedback governs the physics of bacterial motility, growth, and sensing in soil across three scales---the single pore, the mesoscale of many pores, and the broader landscape.

\end{abstract}
\maketitle


\noindent\emph{``We know more about the movement of celestial bodies than about the soil underfoot."} 
\begin{flushright}
--- Attributed to Leonardo Da Vinci
\end{flushright}

\section{INTRODUCTION}
\label{sec:intro}
A bacterium has no eyes, no ears, no brain; it is arguably the simplest
form of life. And yet, collectives of these cells perform complex functions
critical to life. Indeed, most life on this planet would not exist without bacteria in the ground beneath our feet, which number up to a billion cells per gram of soil~\cite{tecon2017}. As these cells decompose organic matter, they release the nutrients that plants need to grow, sustaining the primary production that feeds nearly all life on land. The same metabolism governs the fate of soil carbon, which exceeds that in the atmosphere and all living vegetation combined~\cite{schmidt2011persistence,Lehmann2015,jobbagy2000vertical}. Whether soil releases that carbon or locks it away is decided largely by the activity of the bacteria living inside it~\cite{Schimel2012,Young2004, wilpiszeski2019}; bacteria metabolize the labile\keyterm{Labile}{Organic matter (e.g., simple sugars, amino acids, organic acids) that bacteria decompose.}fraction of soil organic matter---simple sugars, amino acids, and organic acids---whereas carbon stabilized on mineral surfaces or locked in complex molecules turns over far more slowly. Motile bacteria traverse the pores of wet soil following gradients of oxygen and chemicals, such as exudates secreted by plant roots~\cite{tecon2017,Liu2024}.\keyterm{Exudates}{Carbon-rich compounds released by plant roots into the surrounding soil.}The cells eventually colonize the roots and soil grain surfaces, growing as biofilms\keyterm{Biofilm}{A surface-attached bacterial collective encased in self-produced extracellular polymeric substances (EPS).}that secrete enzymes to break down complex organic carbon and release it as CO$_2$~\cite{Schimel2012}. The same cells also drive the nitrogen cycle that governs soil fertility and the greenhouse-gas exchanges between soil and air. Understanding how these collectives move and grow is therefore not only of fundamental interest in biological physics; it bears directly on food security and on how our land responds to a changing climate.

Two features make bacterial life in soil rich but uniquely challenging to study. First, soil is heterogeneous in several ways at once~\cite{tecon2017, vos2013, wilpiszeski2019} \textbf{[Figures~\ref{fig:pore}A, \ref{fig:meso}A, \ref{fig:landscape}A]}. The space between solid soil grains is tortuous and hierarchically structured, with pore sizes ranging from smaller than a micron to as large as millimeters, imposing varying degrees of confinement on the cells. The pore fluid is not a passive solvent: polymers from root exudates and organic matter make it a viscoelastic fluid\keyterm{Viscoelastic fluid}{A fluid that both exhibits viscous flow and stores elastic stress, characterized by a relaxation time $\lambda$ over which stored stress decays.}and, where concentrated (e.g., by drying), a yield-stress gel~\cite{read1999, benard2019, nazari2022, naveed2019}.\keyterm{Yield stress}{The minimum stress a gel must experience before it flows; below it, the material behaves as a soft solid.}Oxygen, dissolved carbon, and root exudate concentrations vary over micron-to-millimeter distances~\cite{watt2006}, exposing cells to steep chemical gradients as they move. Second, the bacteria, in turn, shape this heterogeneous environment across multiple scales. Biofilm-secreted EPS\keyterm{Extracellular Polymeric Substances (EPS)}{Polysaccharides, proteins, extracellular DNA, and other macromolecules secreted by bacteria.}binds grains and clogs pores, altering soil pore structure and, with it, bulk hydraulic conductivity, elasticity, and water retention~\cite{volkBiofilmEffectSoil2016}. Bacterial consumption and enzyme production additionally sculpt the chemical landscape that the cells sense and respond to~\cite{kuzyakovMicrobialHotspotsHot2015a, spohnSpatialTemporalDynamics2014a}. This view of soil as a self-organized, living material is long-standing in soil science~\cite{Young2004, wilpiszeski2019} and is gradually being incorporated in physics-based models~\cite{noronha2025}.

 Nevertheless, most of what we know about the physics of bacterial motility and growth comes from studies of cells in idealized, fixed environments, such as in bulk fluids or at flat surfaces. The theories built to describe bacterial collectives similarly idealize their surroundings. Continuum models of directed motility (Keller-Segel)~\cite{kellersegel1971}, growth-induced spreading (Fisher--KPP)~\cite{fisher1937, kpp1937}, and active matter~\cite{marchetti2013, bechinger2016}, along with recent syntheses of colony growth and form~\cite{porter2025}, take the physical environment as a fixed backdrop, with its geometry and rheology prescribed in advance, acting on the cells but rarely reshaped by them. However, as we review in this article, experiments using transparent mimics~\cite{dijksman2026light} that help isolate soil's complexities---e.g., confinement, rheology, gradients---increasingly reveal how strongly these factors shape bacterial activity, and how the bacteria reshape them in turn. This feedback requires current theoretical frameworks to be extended and new theories to be created, and we highlight opportunities to do so.

This review is organized by examining these bacterial dynamics across three scales: within a single pore (\S\ref{sec:pore}), at the ``mesoscale'' corresponding to many pores (\S\ref{sec:meso}), and over landscapes, where collectives spread across roots and the surrounding terrain (\S\ref{sec:landscape}). Most of the phenomena we discuss have been studied in transparent mimics or other idealized lab settings~\cite{sharmaTransparentSoilMicrocosms2020, downie2014, dijksman2026light, zenglerEcoFAB2019, gaoEcoFABprotocols2018, grossmannRootChip2011, massalhaLiveImaging2017} rather than inside opaque natural soil, where instead only physicochemical or macroscopic observables---e.g., pore structure, water distribution, chemical composition, hydraulic conductivity, mechanical stability, and gas fluxes---are typically accessible. Tracking and probing the biological activity of cells in real, opaque soil remains a central experimental challenge, although progress is being made~\cite{orphanFISHSIMS2001, caroHydrogenStableIsotope2023}. We therefore describe each mechanism in the setting where it has been measured; estimate the governing rates and scales in real soil; distinguish what phenomena have actually been verified in soil from what remains predicted; and suggest measurements that would test them, as well as prospects for new theoretical development. Throughout, we propose that much of this scattered phenomenology can be understood by comparing a characteristic bacterial rate, length, stress, or energy scale against the corresponding one set by soil, defining key dimensionless parameters that organize disparate phenomena across soil conditions. Our focus is complementary to that of quantitative microbial ecology, which has shown that community-level outcomes---e.g., which species coexist, how communities assemble, and how function follows from composition---can be strikingly simple and predictable, often emerging from species interactions even in well-mixed cultures~\cite{goldford2018, dalbello2021, hu2022}. Here, we instead examine the feedback between bacteria and their physical surroundings. We focus on the case of water-saturated soil, because even in this limit, the feedback between bacteria and soil heterogeneity generates a wealth of physics we do not yet understand. Across all three scales a single theme recurs: the richest physics of life in soil lives not in the cells or their habitat alone, but in the feedback between them---a coupling the field is beginning to map, and one of the most inviting frontiers open to biological physics.

\section{THE PORE SCALE}
\label{sec:pore}
\begin{figure*}[t]
\centering
\includegraphics[width=\textwidth]{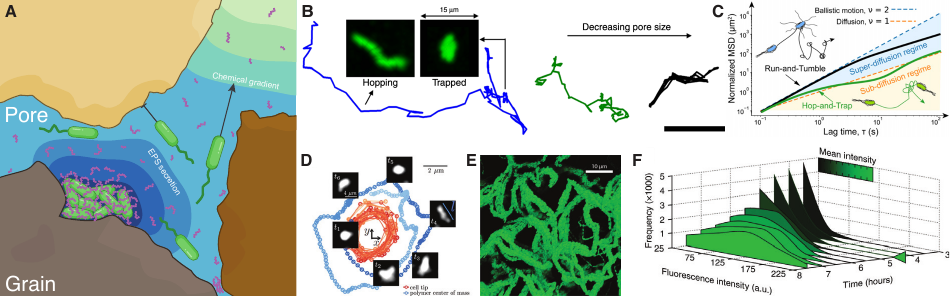}
\caption{\textbf{At the pore scale, confinement and fluid rheology
set how bacterial collectives move, grow, and sense their surroundings.}
(\textbf{A}) Schematic of a single pore: cells swim with runs truncated
by the solid grain walls, bias their motion up a chemical gradient, and
attach to grain surfaces where they secrete EPS to form a biofilm
while shedding EPS into the pore fluid. (\textbf{B}) Experimental trajectories of \emph{E. coli} moving in a porous medium showing that as the pore size decreases, a swimming cell crosses over from
run-and-tumble motion to intermittent ``hopping-and-trapping'', idling
at grain surfaces before reorienting away~\cite{bhattacharjeeBacterialHoppingTrapping2019}. [Scale bar,~$10~\upmu$m.] (\textbf{C}) Schematic of the corresponding single-cell mean square displacement (MSD), showing how transient trapping introduces sub-diffusive motion at intermediate lag times and causes the long-time diffusivity to decrease~\cite{zhangBacterialMotilityPatterns2026}.(\textbf{D}) Experimental visualization showing that a swimming \emph{E. coli} cell strains the surrounding polymer solution,
stretching the polymers and generating elastic stresses that act back on its motion~\cite{patteson2015}. (\textbf{E}) Experimental 3D reconstruction showing that polymer in the pore fluid can hold dividing \emph{E. coli} cells end-to-end, templating elongated, serpentine cell
``cables''~\cite{GonzalezLaCorte2025}. (\textbf{F}) Experimentally-measured distributions of fluorescence intensity reporting the stress response of \emph{E. coli} growing in a confined chamber; green arrow marks the time when the colony has filled the fixed volume, increasing the self-imposed mechanical stress that develops against confinement~\cite{chuSelfinducedMechanicalStress2018}. Panels (B-F) adapted with permission from their respective sources.
}
\label{fig:pore}
\end{figure*}

Soil is a porous medium: mineral grains and organic matter packed into a solid matrix and threaded by a connected void, the pore space. Partitioning this continuous network into discrete ``pores'' is a conceptual division. Nevertheless, it is a useful one, because a single pore is the smallest fundamental unit of the soil habitat: a cavity of fixed size and shape, holding a fluid of a given rheology and bounded by grain walls. Bacterial cells must contend first with these immediate physical surroundings; pore sizes span $\sim10^0-10^2~\upmu$m, so relative to a micron-scale cell the degree of confinement varies over orders of magnitude, which along with the composition of the pore fluid sets the terms on which cells move, grow, and make lifestyle decisions. The same pore that truncates a swimming cell's runs (\S\ref{sec:pore}.1) also screens the flows it generates (\S\ref{sec:pore}.2), holds the polymer-laden fluid it must swim through (\S\ref{sec:pore}.3), templates the colony it grows into (\S\ref{sec:pore}.4), and traps the signals that govern its gene expression (\S\ref{sec:pore}.5) \textbf{[Figure~\ref{fig:pore}A]}. The cells reshape that backdrop in turn---secreting EPS that alters the pore fluid, and growing colonies that narrow or clog the pore---foreshadowing the feedback that dominates at larger scales.

\subsection{Pore geometry reshapes single-cell motility}
In bulk liquid, a cell of the model flagellated bacterium \textit{Escherichia~coli} swims in nearly straight ``runs'' at speed $v\sim30~\upmu$m/s, interrupted by nearly-instantaneous reorientation events (``tumbles'') every $\tau_{r}\sim1~$s~\cite{bergColiMotion2004}. Over longer times, the cell's trajectory is therefore a random walk of step length $\ell_{\text{run}}=v\tau_r\sim30~\upmu$m, so the cell density obeys a diffusion equation with an effective diffusivity $D\approx v\ell_{\text{run}}\sim900~\upmu$m$^2$/s set entirely by the run speed and length; this description applies for other bacteria that similarly swim using run-and-tumble motility. A natural expectation for motility in a porous medium is that the same description applies with $D$ simply rescaled: once the characteristic pore size $\ell_{\text{pore}}$ drops below the run length $\ell_{\text{run}}$, collisions with grain surfaces simply truncate runs, lowering $D$ to $\approx v\ell_{\text{pore}}$ instead. However, bulk transport measurements show that this picture fails for soil-like media: it overpredicts the measured diffusivity, with the discrepancy growing in finer packings with smaller pores~\cite{fordSmith2011,bartonFord1995,sherwood2003,orSmets2007, fordHarvey2007}.

Single-cell imaging in transparent soil mimics has since resolved why. As the pore size $\ell_{\text{pore}}$ decreases below the run length $\ell_{\text{run}}$, a swimming cell increasingly meets a grain surface where hydrodynamics forces the cell to idle before reorienting away. In strongly confining pores where $\ell_{\text{pore}}$ approaches the cell body size $\ell_{\text{cell}}$, steric hindrance also increasingly dominates~\cite{spagnolieSwimmingComplexFluids2023, gaoSelectiveTrappingBacteria2026}, causing these idling events to lengthen into longer ``trapping'' events that can last as long as $\tau_t\sim10$~s. The cell then ``hops'' through the pore space over a step length $\sim\ell_{\text{pore}}$ before it is transiently trapped again \textbf{[Figure~\ref{fig:pore}B]}, leading to an effective long-time diffusivity $D\approx\ell_{\text{pore}}^2/\tau_t$ whose value is now set by the competition between cellular activity and pore-scale confinement \textbf{[Figure~\ref{fig:pore}C]}~\cite{bhattacharjeeBacterialHoppingTrapping2019,Bhattacharjee2021porous,zhangBacterialMotilityPatterns2026}. Beyond model lab environments, signatures of this hopping-and-trapping motility have been observed in biological tissues~\cite{akolpoglu2022magnetically}.

These observations provide clues for constructing a predictive theory of bacterial motility in soil; in particular, they suggest that the diffusivity $D$ follows from the pore-size distribution $P(\ell_{\text{pore}})$ and cell reorientation statistics. Simulations and model experiments support this picture: the long-time diffusivity collapses onto a single geometric length $\sim\min(\ell_{\text{run}},\ell_{\text{pore}})$ across media of varying pore size, shape, porosity, and disorder~\cite{kurzthalerGeometricCriterion2021, pietrangeliUniversalLawDispersal2025, mattinglyCoarsegrainingBacterial2025, dehkharghaniSelftransport2023}. The remaining step is to carry these idealized results to the more complex three-dimensional (3D) space of real soil. The geometry dependence already yields a clean expectation, captured by the confinement ratio $\ell_{\text{run}}/\ell_{\text{pore}}$ that compares the bulk run length to the pore size~\cite{kurzthalerGeometricCriterion2021,spagnolieSwimmingComplexFluids2023}: motility should be trap-dominated where $\ell_{\text{run}}/\ell_{\text{pore}}\gtrsim1$, as in fine-textured silt ($\ell_{\text{pore}}\sim1-20~\upmu$m, $\ell_{\text{run}}/\ell_{\text{pore}}\sim2-30$), and cross over to conventional run-and-tumble where $\ell_{\text{run}}/\ell_{\text{pore}}\lesssim1$, as in coarse sand ($\ell_{\text{pore}}$ approaching $\sim400~\upmu$m, $\ell_{\text{run}}/\ell_{\text{pore}}\sim0.1$). Moreover, because trapping is dominated by the smallest pores, we expect that two soils with the same mean pore size but different polydispersity should spread cells at measurably different rates. In nature, however, this picture is additionally complicated by variations in water saturation: as soil dries, air invades the pores and forces water to thin to films coating the grain surfaces, which can suppress swimming altogether, restricting motility to brief wet episodes~\cite{dechesne2010}.

Pore-scale confinement also reshapes how a cell biases its random walk in response to a chemical gradient---a process known as chemotaxis.\keyterm{Chemotaxis}{Biased movement up or down a chemical gradient, achieved by modulating the statistics of an underlying random walk rather than steering directly.}In bulk liquid, \textit{E.~coli} bias their motion primarily by modulating run \emph{duration}, lengthening runs that move up an attractant gradient, with a secondary mechanism that biases the \emph{orientation} of the next run~\cite{saragosti2011}. What is the purpose of this second, apparently redundant mechanism? Confinement resolves this puzzle. In a tight pore space, hop lengths are constrained by pore geometry and can no longer be biased; direct measurements in transparent soil mimics show that \textit{E.~coli} then bias their motion primarily using orientational bias instead~\cite{bhattacharjee2021}---an effect confirmed for the soil bacterium \textit{Pseudomonas putida} as well~\cite{beierDecipheringDualChemotaxis2026}. The apparently secondary mechanism is therefore what sustains chemotaxis under confinement, so carrying more than one bias mechanism lets a cell perform chemotaxis across the full range of settings it encounters in soil. A predictive theory should again derive these strategies from geometry, taking $P(\ell_{\text{pore}})$ and the per-mechanism reorientation statistics to predict which bias dominates, and the chemotactic drift it yields, as the pore size varies. Why a cell maintains multiple mechanisms to bias its motion at all is then an open evolutionary question: if orientational bias is selected because it preserves chemotaxis under the confinement that disables run-length bias, a species' mix of strategies should track the pore sizes of its native habitat, which can be tested by comparing soil bacteria from coarse and fine textures.

\subsection{Hydrodynamics in the pore space}
A swimming bacterial cell both generates fluid flow and, in saturated soil, is bathed in the ambient flow of water through the pore space. In this confined space, the soil grains screen the cell-generated flow, damping its hydrodynamic coupling to other cells, and the ambient flow through pores imposes shear stresses that can reorient the cell. The broader hydrodynamics of swimming near surfaces and in confined channels have been reviewed in depth elsewhere~\cite{conradPolingSkutvik2018}; here we discuss only these two soil-specific effects.

Hydrodynamic interactions between cells are known to generate fascinating collective behaviors in bulk dense suspensions, such as large-scale active turbulence~\cite{Wioland2013} and boosted transport of nutrients, waste, and chemical signals~\cite{Xu2019}. However, we do not expect that they arise in soil. The flow a swimming cell generates decays approximately as a force dipole, but in a soil-like packing of permeability $k$, the solid grain surfaces screen this flow beyond the Brinkman length $\ell_\text{Br}\sim\sqrt{k}$.\keyterm{Brinkman length}{The distance over which a porous matrix screens fluid flow, $\ell_\mathrm{Br}\sim\sqrt{k}$ with $k$ the permeability; beyond it, hydrodynamic interactions between cells are cut off.} Measured saturated soil permeabilities span $k\sim10^{-2}-10^{2}~\upmu\textrm{m}^2$ across textures~\cite{freezeCherry1979}, corresponding to $\ell_\mathrm{Br}\sim10^{-1}-10^{1}~\upmu\mathrm{m}$. Hence, the long-range hydrodynamic interactions between cells are screened once $\ell_\text{Br}$ falls below their separation. Capturing these effects theoretically requires an active-suspension description in which the fluid is not free, but screened by the porous matrix; continuum models of swimmer suspensions in an idealized porous medium are a step in that direction~\cite{almoteriBrinkmanCollective2025}, but connecting them to the disordered pore geometry of real soil remains open.

When water flows through the pore space (e.g., due to gravity-driven infiltration, drying, or root-driven uptake), it imposes shear at a rate $\dot\gamma\approx u/\ell_{\text{pore}}$, where $u\sim 0.1-100~\upmu\textrm{m}/\textrm{s}$ is the interstitial flow speed across different soil textures~\cite{hillelEnvironmentalSoilPhysics2009,rooseFowlerDarrah2001}. Since the coarsest soils combine the fastest flow with the widest pores, natural soils experience $\dot\gamma\sim0.1-1~\mathrm{s}^{-1}$. Because this rate exceeds that at which the cell's swimming direction randomizes, $D_r\sim0.1~\mathrm{s}^{-1}$ for \textit{E.~coli}~\cite{drescher2011fluid}, we expect that during flow elongated cells align with the streamlines, accumulate at pore walls, and can swim upstream~\cite{Rusconi2014,torres2024enhancement}, particularly in coarser soil textures. This competition is quantified by the shear P\'eclet number, $\Pe_{\text{shear}}\equiv\dot\gamma/D_r\sim1-10$, with flow-induced alignment arising when $\Pe_{\text{shear}}\gtrsim1$. This effect should be intermittent in natural soil undergoing cyclic wetting and drying, however: the water flux falls toward zero between wetting events, so the alignment should track the hydrograph---with wall accumulation peaking during and shortly after rain, irrigation, or active transpiration, and decaying as the soil drains or a plant root's demand subsides.

\subsection{Swimming through the pore fluid: from viscoelastic to yield-stress}
The pore fluid is not a simple solvent. Root exudates and microbial EPS make it a polymer solution, which can be a viscoelastic fluid when more dilute, and a yield-stress gel when more concentrated (e.g., by drying). Because the cells themselves secrete the EPS, they tune the rheology of the very fluid they must then swim through---a pore-scale feedback whose larger-scale consequences for soil structure we revisit in \S\ref{sec:landscape}~\cite{robersonFirestone1992, costa2018}. The general physics of swimming in complex fluids has been reviewed in depth elsewhere~\cite{spagnolieSwimmingComplexFluids2023}; here we focus on the two regimes of swimming in a viscoelastic fluid and a yield-stress gel as they relate to life in soil.

In the viscoelastic fluid case, extracellular polymers can change bacterial swimming in ways viscosity alone does not capture. At lower polymer concentrations, the cell's own flow can be strong enough to stretch polymer molecules, generating elastic stresses that act back on it \textbf{[Figure~\ref{fig:pore}D]}---suppressing tumbling and lengthening runs, and increasing the swimming speed~\cite{patteson2015}. In particular, this effect arises because the swimming cell strains the fluid at a rate $\dot\gamma_c\sim v/\ell_{\text{cell}}\sim30~\mathrm{s}^{-1}$ exceeding the rate at which the constituent polymers relax, which can be as small as $\lambda^{-1}\sim0.1-10~\mathrm{s}^{-1}$ for high-molecular weight polymers, where $\lambda$ is the relaxation time~\cite{spagnolieSwimmingComplexFluids2023}. This competition is quantified by the Weissenberg number, $\Wi\equiv\dot\gamma_c\lambda$, with swimming-induced polymer stretching, and the resulting enhancement in swimming speed, arising when $\Wi\gtrsim1$; we estimate that $\Wi$ can be as large as $\sim3-300$ for high molecular weight exudates and EPS, although $\lambda$ has not been measured for these soil polymers. This enhancement reverses at higher polymer concentrations due to a polymer-depleted layer near the cell body~\cite{martinez2014flagellated}; the same behavior appears in colloidal suspensions, indicating that it is a more generic property of swimming in complex fluids~\cite{kamdarCheng2022}. A second, slower comparison is set by the Deborah number $\De\equiv\lambda/\tau_r$, which compares the polymer relaxation time to the interval $\tau_r\sim\min(\ell_{\text{run}},\ell_{\text{pore}})/v\sim0.1-1~\mathrm{s}$ between cell reorientations in the confined pore space; when $\De\gtrsim1$, which we expect arises in concentrated soil pore fluid, the polymeric fluid carries a memory of one run into the next, which can feed back on collective behavior. For example, it could suppress the self-trapping behind motility-induced phase separation (MIPS) when the two time scales are comparable~\cite{diterlizzi2026}, although extending MIPS theory from a Markovian bath to a viscoelastic one remains an open theoretical problem.

As the polymer in the pore fluid concentrates---e.g., due to drying, or near EPS-producing biofilms or root surfaces---it becomes a gel that resists motion until a threshold yield stress $\sigma_y$ is exceeded. Exudate and microbial EPS gels can have $\sigma_y\sim10^{-1}-10^3$~Pa~\cite{naveed2019,pavlovsky2013}. A bacterial cell must then generate enough torque to yield this gel before its flagellum can turn, and enough thrust to displace the yielded material before it can advance, as verified for model swimmers~\cite{nazariHelicalLocomotionYield2023}. Whether the cell advances is set by the Bingham number $\Bi\equiv\sigma_y/\sigma_\mathrm{prop}$, which compares the gel yield stress to the propulsive stress $\sigma_\mathrm{prop}\approx6\pi\mu v/\ell_{\text{cell}}\sim1~\mathrm{Pa}$~\cite{chattopadhyaySwimmingEfficiencyBacterium2006a} the cell exerts when swimming through pore water of viscosity $\mu\sim1~\mathrm{mPa}\cdot\mathrm{s}$. Because $\sigma_\mathrm{prop}$ lies in the lower portion of the $\sigma_y\sim10^{-1}-10^3~\mathrm{Pa}$ range, $\Bi<1$ in dilute pore fluid and the cell swims freely, but $\Bi$ rises past unity and the cell is immobilized as drying or EPS raises the local yield stress---whereupon it continues to grow, divide, and form a new colony, as discussed next.

\subsection{Collective growth is shaped by polymers, crowding, and porous confinement}
When bacterial cells grow rather than swim, the surrounding polymer and confinement template the shape the resulting colony takes---as governed by the competition between a stress that organizes the collective and one that resists it. For example, a polymer-rich pore fluid can generate a depletion attraction\keyterm{Depletion attraction}{An effective attraction between bodies in a polymer solution, driven by the osmotic pressure of polymers excluded from the gap between them.} holding dividing cells end-to-end, causing them to grow into long, multicellular, serpentine ``cables'' \textbf{[Figure~\ref{fig:pore}E]}~\cite{GonzalezLaCorte2025}. This phenomenon occurs when the ratio $U_d/k_BT$, which compares the depletion attraction energy $U_d\sim\Pi_pV_o$ to the thermal energy $k_BT$, exceeds unity; here, $\Pi_p\sim nk_BT$ is the polymer osmotic pressure, where $n$ is the polymer number density, and $V_o\sim10^{-3}-10^{-1}~\upmu\textrm{m}^{3}$ is the excluded-volume overlap between adjacent micron-scale cells. The underlying attraction is well documented using natural polysaccharides~\cite{dorken2012depletion, schwarzlinek2010polymer, secor2018entropically, niu2021depletion}; the distinct serpentine cable form emerges when the cells also grow and divide while held end-to-end, as demonstrated so far in synthetic polymer and gut mucus~\cite{GonzalezLaCorte2025}. Root exudate and microbial EPS are concentrated polymer solutions and should act as depletants, with an estimated $n\approx 10^{2}-10^{3}~\upmu\mathrm{m}^{-3}$~\cite{or2007} and therefore $U_d$ can become as large as $\sim 10k_BT$, so we expect that cables can form wherever the local polymer concentration pushes the depletion energy above the thermal scale.

Another organizing mechanism is cellular alignment rather than attraction. A solution of elongated subunits, such as large polymers or rod-like molecules, can spontaneously align along a common axis once they are crowded enough, forming a nematic liquid crystalline phase: a state with orientational order, in which the subunits share an average direction (the director)\keyterm{Director}{The local average orientation axis of a nematic phase---the direction along which rod-like subunits align.} without positional order. Indeed, bacterial EPS can exhibit nematic ordering~\cite{repula2022biotropic}. If root exudate and soil microbial EPS likewise form nematic phases at natural concentrations, a cell that misaligns from the local director pays an elastic energy penalty $U_{LC}\sim W\ell_{\text{cell}}^2$~\cite{rapini1969distorsion}, where the parameter $W$ describes how strongly the surrounding nematic is anchored to the cell surface. In a nematic liquid crystal, $W\sim10^{-7}-10^{-4}~\mathrm{J}/\mathrm{m}^2$, so the ratio $U_{LC}/k_BT$ can reach $\sim10^{4}$; as with the depletion interaction, once the organizing energy $U_{LC}$ exceeds $k_BT$ the dividing cells are held along the local director and forced to grow end-to-end as aligned ``chains''~\cite{corteMorphogenesisBacterialColonies2025}---potentially providing bacteria a new way to colonize new pores in soil.

Even without an external director, growing bacteria can generate their own order: dividing cells push neighbors into local nematic alignment, organizing a non-motile colony into microdomains of aligned cells separated by topological defects~\cite{youGeometryMechanicsMicrodomains2018, dellarcipreteGrowingBacterialColony2018, Doostmohammadi2018}.\keyterm{Topological defect}{A point or line where orientational order breaks down; in growing colonies the motile $+\tfrac12$ defects mark where new cell layers form.} The motile $+\tfrac12$ defects then set where the colony escapes into the third dimension: new cell layers form preferentially at $+\tfrac12$ defects, as shown directly for the soil bacterium \textit{Myxococcus xanthus} and captured by an extensile active-nematic model, while cells verticalize through localized mechanical instabilities triggered by division~\cite{copenhagenTopologicalDefectsLayer2021, berozVerticalizationBacterialBiofilms2018}. The microdomain size $\ell_\mathrm{nem}$ is set by the competition between orientational stiffness and extensile growth stress, so nematic order persists only on scales below that length; while it is still challenging to estimate the values of these competing stresses, lab measurements indicate that $\ell_\mathrm{nem}\sim10-20~\upmu$m~\cite{youGeometryMechanicsMicrodomains2018, you2021confinement,Doostmohammadi2018}. In 3D, confinement of growing bacterial cells within a gel selects the form of ordering through the stiffness ratio $G_{\text{gel}}/G_{\text{colony}}$, which compares the elastic modulus of the surrounding gel to that of the colony: when $G_{\text{gel}}/G_{\text{colony}}>1$ the constituent cells are forced to order into a bipolar, anisotropic structure, whereas when $G_{\text{gel}}/G_{\text{colony}}<1$ the colony remains disordered and isotropic~\cite{zhangMorphogenesisCellOrdering2021}. Concentrated root exudates and secreted microbial EPS can have $G_{\text{gel}}\sim10^{-1}-10^3$~Pa~\cite{naveed2019,pavlovsky2013}, comparable to or potentially exceeding that of a growing colony, $G_{\text{colony}}\sim10^1-10^3~$Pa~\cite{pavlovsky2013,charltonViscoelasticityBiofilms2019,ohmuraMicrorheologyBiofilms2024}; therefore, growing bacterial colonies should stay isotropic in pores with dilute organic content, but order into anisotropic shapes as drying and root exudate production concentrate polymers in the pore space. Whatever shape it takes, a colony that fills a pore narrows or clogs it, reshaping the very confinement that set its form---a pore-scale feedback that, compounded across many pores, controls how biofilms restructure flow and hydraulic conductivity (\S\ref{sec:meso}, \S\ref{sec:landscape})~\cite{thullner2002, volkBiofilmEffectSoil2016}.

\subsection{Sensing and lifestyle commitment under confinement}
As a soil bacterium interacts with its surroundings, environmental stimuli can cause it to switch between multiple lifestyles:\keyterm{Planktonic versus sessile}{Free-swimming, dispersing versus growing, surface-attached cells.}for example, it can freely swim through water-filled pores searching for a root or a nutrient patch (planktonic), or it can choose to attach to a grain or root surface, secrete an EPS matrix, and grow in a surface-attached (sessile) biofilm instead. Other lifestyle switches include secreting extracellular enzymes or sporulating under starvation. One environmental stimulus that drives this switching is mechanical: as a confined collective grows in a pore, the mechanical stress that develops as it presses against the surrounding grains can upregulate stress response and trigger the transition to biofilm formation \textbf{[Figure~\ref{fig:pore}F]}~\cite{chuSelfinducedMechanicalStress2018}. Another stimulus is chemical, which can drive lifestyle switching through a process called quorum sensing:\keyterm{Quorum sensing}{Cell--cell communication in which bacteria secrete and detect a diffusible autoinducer, switching gene expression once its local concentration crosses a threshold.} cells continuously secrete and detect a small diffusible signal, the autoinducer, and once its local concentration crosses a threshold it triggers a coordinated change in gene expression~\cite{perezHeterogeneousResponseQuorumSensing2010}. This threshold is often thought to be a reporter of local population density, since more cells make more signal~\cite{morenogamezQuorumSensingWisdom2023}. However, pore-scale confinement changes this paradigm. The autoinducer concentration $a_{\mathrm{ai}}$ a single cell maintains against diffusive loss (with diffusivity $D_\mathrm{ai}\sim10^{3}~\upmu\mathrm{m}^2/\mathrm{s}$) is $\sim p/(4\pi D_\mathrm{ai} \ell_{\text{cell}})$ for a cell of size $\ell_{\text{cell}}$ emitting autoinducer at a rate of $p$ molecules per unit time~\cite{bergpurcell1977}. In a water-filled pore of size $\ell_{\text{pore}}\sim1-100~\upmu$m, the autoinducer equilibration time is only $\tau_\mathrm{ai}\sim \ell_{\text{pore}}^2/D_\mathrm{ai}\sim10^{-3}-10^1$~s, so even one to a few confined cells can rapidly sustain a steady-state concentration above the quorum-sensing threshold~\cite{boedickerMicrofluidicConfinementSingle2009, Carnes2010}. Indeed, once diffusive loss is suppressed by confinement in this way over longer times, the autoinducer instead accumulates as cells produce it, and a collective of cell density $\rho$ reaches the quorum-sensing threshold $a_{\mathrm{ai}}^*$ over a buildup time $\tau_\mathrm{qs}\sim a_{\mathrm{ai}}^* /(\rho p)$. We therefore expect that in soil, quorum sensing-mediated lifestyle switching is far more prevalent than in bulk systems, set by the local geometry, not just local population density---particularly in the smallest pores of soil, as we examine further in \S3.1.

The direction of the switch between the planktonic and sessile lifestyles is not fixed. Depending on the bacterial species and signal transduction circuit, a rising autoinducer level can force cells to switch from planktonic dispersal to sessile growth in a biofilm, or, conversely, repress biofilm formation and favor being in the motile state---for example, helping mature biofilms actively disperse cells outward when local conditions become unfavorable. Lifestyle switching is therefore bidirectional, and which way it points is set by cellular regulatory wiring as much as by cell density~\cite{morenogamezQuorumSensingWisdom2023}. Flow shifts the balance further, as quantified by a P\'eclet number $\Pe_{\text{flow}}\equiv u\ell_{\text{pore}}/D_\mathrm{ai}$ comparing advective removal of autoinducer at pore-flow speed $u$ to its diffusion: where $\Pe_{\text{flow}}>1$ the pore flow sweeps autoinducer away faster than it accumulates, so the quorum-sensing response is likely suppressed on flow-exposed surfaces and proceeds only in sheltered grooves, dead ends, and slow-flow pore interiors where $\Pe_{\text{flow}}<1$. Thus, genetically identical cells could adopt different states depending on their location in the pore space itself~\cite{Kim2016, Emge2016, mok2019geometric}. Clarifying how these behaviors play out across the complex pore space of soil and organize across the root zone as a whole is an important direction for future research.

Beyond autoinducers, bacterial cells sense and respond to other physicochemical factors in their surroundings, such as changes in nutrient and water availability, or exposure to stressors. In many cases, they respond not just to the absolute level of a stimulus, but how fast it changes, characterized by the time scale $\tau_{\text{sensing}}$~\cite{youngRateEnvironmentalChange2013}; when a stimulus crosses its threshold value, it not only activates the cell's response, but switches on intracellular feedback that suppresses this response after a time $\tau_{\text{feedback}}\sim10^2-10^3~\mathrm{s}$, although the exact value depends on the specific bacterial biochemical circuit~~\cite{segall1986temporal,youngRateEnvironmentalChange2013}. Whether bacteria respond to a stimulus is therefore set by the ratio $\tau_{\text{feedback}}/\tau_{\text{sensing}}$; when this ratio exceeds unity i.e., when the stimulus varies faster than $\tau_{\text{feedback}}^{-1}$ does it elicit an appreciable response, buffering the influence of slower variations in the environment. In the context of soil, this comparison of time scales suggests that abrupt wetting/rewetting events and sudden nutrient pulses created by e.g., cell lysis or introduction of new dead organic matter ($\tau_{\text{sensing}}\sim10^0-10^2~\mathrm{s}$) influence microbial activity more than slower diurnal changes in exudation, temperature, and moisture ($\tau_{\text{sensing}}\gtrsim10^5~\mathrm{s}$), which are instead filtered out.

\section{THE MESOSCALE}
\label{sec:meso}
\begin{figure*}[t]
\centering
\includegraphics[width=\textwidth]{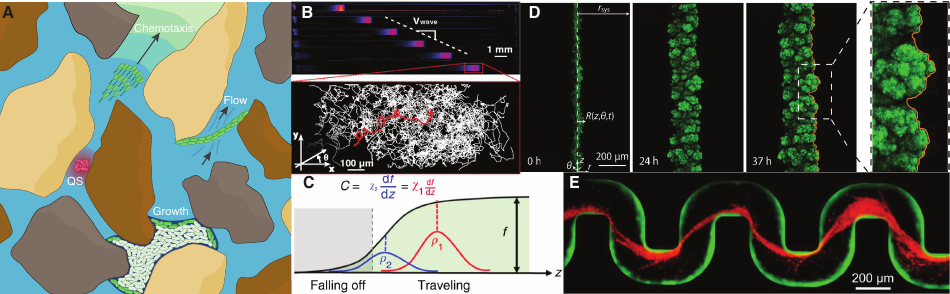}
\caption{\textbf{At the mesoscale, bacterial collectives shape and respond to the chemical fields they create as they migrate, grow, and switch between lifestyles.}
(\textbf{A}) Schematic across many pores: collectives migrate as
chemotactic fronts up chemical gradients, form streamers in flow,
upregulate quorum sensing (QS) in tight pores (red), and grow as dense colonies that
push on the surrounding grains, with metabolic activity confined to
an active surface layer (green).
(\textbf{B}) Top panel shows a series of fluorescent micrographs (time progressing from top to bottom) of a chemotactic front of \emph{E. coli} advancing up a self-generated gradient in a microfluidic channel; lower panel shows individual cell trajectories~\cite{saragosti2011}.
(\textbf{C}) Schematic of the self-sorting mechanism in a front migrating along $z$~\cite{fu2018}: cells sense the chemoattractant as a
perceived signal $f(z)$ (black) whose local steepness $df/dz$ is
greatest at the back of the front and shallower at its leading edge. Every
bacterial cell that travels with the band moves at the same collective speed
$C=\chi_i\,(df/dz)$, so two cells with chemotactic coefficients
$\chi_1>\chi_2$ occupy different positions: the stronger climbers
($\rho_1$, red) follow the shallow-gradient leading edge, the weaker
($\rho_2$, blue) the steep-gradient back. A cell whose chemotactic
coefficient is too low to match $C$ at any position falls off the back
and is left behind.
(\textbf{D}) Confocal micrographs showing that a dense colony of \emph{E. coli} growing on abundant nutrient develops a fractal,
``broccoli''-like morphology through a growth-driven surface
instability~\cite{martinezcalvo2022}.
(\textbf{E}) A biofilm streamer of \emph{P. aeruginosa} (red) forms in a microfluidic channel as flow
draws yielded biomass downstream~\cite{drescher2013}; green shows biofilm pre-grown on the channel walls. Panels (B-E) adapted with permission from their sources.}
\label{fig:meso}
\end{figure*}

Across multiple pores ($\sim10^2-10^3~\upmu$m), cells couple to one another by consuming and secreting chemical signals into the shared pore fluid: they deplete nutrient into a gradient they then move along, grow into colonies that their own metabolism sculpts, and decide to switch between these different lifestyles via e.g., quorum sensing \textbf{[Figure~\ref{fig:meso}A]}. As in the previous section, these mesoscale dynamics can be understood through a recurring competition between a length, time, or stress scale characterizing the two-way feedback between the collective and the environment it has itself altered.

\subsection{A bacterial collective shapes the chemical gradients it moves along}
Bacteria bias their motion in response to chemoeffectors, molecules that they either move towards (``attractants'') or away from (``repellents''). This single-cell strategy allows bacterial collectives to navigate complex terrain as coordinated groups: by consuming a chemoeffector, a collective establishes a local gradient that the cells move along as a coherent front, continuing to propagate the gradient through their consumption as they move \textbf{[Figure~\ref{fig:meso}B]}~\cite{adler1966,saragosti2011}. These self-sustaining fronts move at a constant velocity, causing the cells to far outpace their undirected diffusion and enabling them to find new nutrient sources and colonize new terrain~\cite{cremer2019, narla2021, keegstra2022}. In particular, for a single attractant (chemical concentration $a$) and in the absence of growth, the front advances at a speed set by the cells' chemotactic drift velocity, $v_c\approx\chi\nabla f(a)$, where $f(a)$ describes their logarithmic sensing of the attractant~\cite{kalinin2009,fu2018,cremer2019}; the chemotactic coefficient $\chi$ describes their ability to move up the sensed gradient. This form of collective migration has been directly observed in transparent soil mimics~\cite{bhattacharjee2021, sharmaTransparentSoilMicrocosms2020,alharraqSoilTextureRegulates2026}. These experiments indicate that, as with the undirected diffusivity $D$, pore-scale confinement suppresses $\chi$ as well, slowing front migration~\cite{crozeMigrationChemotacticBacteria2011,bhattacharjee2021}---though, unlike for $D$, no prediction of $\chi$ from the pore-size distribution yet exists. Because both coefficients are set by the same confinement-truncated random walk described in \S\ref{sec:pore}.1, a useful next step will be to map the pore-size distribution and reorientation statistics---including the single cell's shift toward orientational rather than run length bias under increasing confinement---onto $\chi$ and $D$, as well as the overall properties of the migrating front.

These fronts are remarkably coherent, even when confronted with multiple forms of heterogeneity. For example, the collective is robust to heterogeneity in cell behavior within itself through self-sorting: cells that are more sensitive to the gradient migrate along its shallower leading edge, while those that are less sensitive trail along the steeper rear, enabling both to move at the same drift velocity and travel as one front \textbf{[Figure~\ref{fig:meso}C]}~\cite{fu2018}. Moreover, because the tactic response weakens as attractant concentration rises, variation in sensing along the front smooths large-scale perturbations in its geometry that may arise from environmental heterogeneity~\cite{bhattacharjee2022, alert2022}. However, not all bacteria direct their motion the same way; several soil and marine bacteria modulate swimming \emph{speed} rather than biasing their run length or orientation---a strategy known as chemokinesis~\cite{son2016, alirezaeizanjaniChemotaxisStrategiesBacteria2020}. Investigating if, and how, such collective behaviors are altered for chemokinetic bacteria is a fascinating direction for future research.

Bacterial growth can boost or compete with chemotactic front migration. When faced with a distinct attractant and nutrient, chemotaxis up the self-generated attractant gradient (ambient concentration $a_\infty$, diffusivity $D_a$) pulls the front forward while growth on the nutrient (at rate $\alpha$) fills in behind it~\cite{cremer2019, narla2021, keegstra2022}. When instead the attractant \emph{is} the nutrient, cells consume (at a per-cell rate $\kappa$) and chase a single resource, and the collective's spreading shifts between two modes as the balance of growth and chemotaxis changes. In this case, the front migrates up a chemoeffector gradient of width $\ell_a\sim\sqrt{D_a\,\tau_\mathrm{dep}}$, where $\tau_\mathrm{dep}\sim a_\infty/(\rho\kappa)$ is the time the collective (cell density $\rho$) takes to deplete the local nutrient, and the time for a cell to chemotax across $\ell_a$ is $\tau_\mathrm{chemo}\sim\ell_a^2/\chi$. Which migration mode dominates is set by the ratio $\tau_\mathrm{chemo}/\tau_\mathrm{grow}$, where $\tau_\mathrm{grow}\equiv\alpha^{-1}$ is the doubling time. When $\tau_\mathrm{chemo}/\tau_\mathrm{grow}>1$ the fast chemotactic front transitions to a slower, growth-driven front: when the cells remain motile, this growth-driven front is a Fisher-KPP wave\keyterm{Fisher-KPP wave}{A constant-speed traveling front arising from the coupling of diffusion and logistic growth.} with speed $v_c\sim\sqrt{D\alpha}$, while if the cells are sufficiently confined that they jam ($D,\chi\to0$), the front advances purely by proliferation pushing cells outward~\cite{farrellMechanicallyDriven2013,amchinInfluenceConfinementSpreading2022}. This crossover is therefore likely to be soil texture dependent: finer soil confines the cells more, suppressing $\chi$ (and the active diffusivity $D$ alongside it; \S\ref{sec:pore}) and tipping the population from a chemotactic front toward a growth-driven, jammed front~\cite{amchinInfluenceConfinementSpreading2022}. Whether confinement drives the same chemotactic-to-growth-driven transition when attractant and nutrient differ remains to be investigated.

Soil supplies more than enough labile chemoeffectors for cells to sense: its pore water contains a diverse array of chemoeffectors supplied by root exudation, cell lysis, and organic-matter breakdown~\cite{deweertFlagellaDrivenChemotaxisExudate2002, rudrappaRootSecretedMalicAcid2008, zhalninaDynamicRootExudate2018}, including free amino acids ($a\sim1-20~\upmu$M), as well as sugars and organic acids ($a\sim0.1-1$~mM near the root surface, decaying to $\sim1~\upmu$M further away)~\cite{jonesCarbonFlowRhizosphere2009, kuzyakovMicrobialHotspotsHot2015a}. While measurements on soil bacteria are sparse, experiments on model species suggest that bacteria can detect and respond to chemoeffectors across a broad range of $a\sim10^{-3}-10^{2}~\upmu$M~\cite{mesibov1973, kalinin2009}---indicating that concentrations in soil fall well within the cell's dynamic range, saturating it at the richest near-root sources. Hence, bacterial chemotaxis in soil is not chemoeffector concentration-limited. Instead, the limiting factor is the chemoeffector gradient strength. A source whose concentration decays over a length $L$ presents a relative gradient $|\nabla a|/a\sim1/L$; across soil's sources, from $L\sim10^2~\upmu$m patches around lysed cells and fresh detritus to $L\sim10^4~\upmu$m root exudate fields, the relative gradient spans $10^{-4}-10^{-2}~\upmu\textrm{m}^{-1}$, reaching or exceeding the range bacteria are tuned to sense near the steepest sources and falling within it at the broader root-exudate scale~\cite{clausznitzer2014}. Only in the dilute region between sources, where the chemoeffector pool is sparse, does the relative gradient fall below the sensing window. We therefore expect that bacterial chemotaxis in soil is gradient-limited: effective in the neighborhoods of roots, fresh detritus, and lysed cells, and weak in the region between them. A second precondition is cell density: for a collective of cells to appreciably shape and subsequently respond to the chemoeffector gradient, a sufficient density of cells ($\gtrsim10^8$~cells/mL) needs to be within the front~\cite{keegstra2022}. Field measurements suggest that soil bacteria could indeed reach densities $\rho\sim10^8-10^9$~cells/mL near chemoeffector hotspots~\cite{Raynaud2014}, meeting this requirement---although these measurements aggregate over all taxa and thus provide an upper bound.

The same dense hotspots are where a collective may switch to forming a sessile biofilm via quorum sensing instead of continuing to disperse, set by which of three characteristic time scales is shortest: the autoinducer-buildup time $\tau_\mathrm{qs}\sim a_{\mathrm{ai}}^* /(\rho p)$ introduced in \S2.5, the nutrient-depletion time $\tau_\mathrm{dep}\sim a_\infty/(\rho\kappa)$, and the dispersal time $\tau_\mathrm{disp}\sim L/v_c$ for cells to chemotax out of a patch of size $L$. If the autoinducer reaches the quorum-sensing threshold first ($\tau_\mathrm{qs}<\tau_\mathrm{dep},\tau_\mathrm{disp}$) the collective switches to growing as a biofilm. By contrast, if nutrient is depleted first ($\tau_\mathrm{dep}<\tau_\mathrm{qs}$), the cells starve before a quorum forms, although in some cases starvation can lower the threshold for quorum sensing~\cite{alcalde2026phosphorus}. If the cells chemotax away first ($\tau_\mathrm{disp}<\tau_\mathrm{qs}$), the autoinducer producers disperse before the quorum sensing signal accumulates, favoring continued migration~\cite{mooreott2022}. At hotspots in soil, we estimate that nutrient is locally depleted within $\tau_\mathrm{dep}\sim10^3-10^5$~s, taking $\kappa\sim10^{-18}-10^{-17}$~mol/cell/s for the actively-metabolizing cells expected at a hotspot~\cite{vinolas2001}; most soil cells are dormant and consume far more slowly~\cite{blagodatskayaKuzyakov2013}. By comparison, we expect that cells chemotactically disperse from patches over $\tau_\mathrm{disp}\sim10^1-10^3$~s (assuming $v_c\sim1-10~\upmu$m/s), and the buildup time $\tau_\mathrm{qs}$ can fall in the same $10^1-10^3$~s range, depending on the autoinducer production rate $p$, the threshold $a^*_\mathrm{ai}$, and signal turnover in soil. Nutrient depletion is therefore typically the slowest of the three processes, so a collective rarely starves before it either commits to forming a biofilm or disperses. The decisive competition is between autoinducer buildup and chemotactic escape ($\tau_\mathrm{qs}$ versus $\tau_\mathrm{disp}$), and because $\tau_\mathrm{qs}$ and $\tau_\mathrm{disp}$ overlap over that same $\sim10^1-10^3$~s range, small differences in local cell density, signal turnover, and gradient steepness can tip cells near a hotspot toward biofilm formation ($\tau_\mathrm{qs}/\tau_\mathrm{disp}<1$) or continued migration ($\tau_\mathrm{qs}/\tau_\mathrm{disp}>1$).

Much of the picture above carries over to the case of oxygen as a chemoeffector, since oxygen is the terminal electron acceptor for aerobic\keyterm{Aerobic bacteria}{Bacteria that use molecular oxygen as the terminal electron acceptor for respiration to produce energy, and therefore require oxygen to live, grow, and carry out metabolic processes.}metabolism. In saturated soil, oxygen reaches bacteria only by aqueous diffusion (with diffusivity $\sim10^3~\upmu$m$^2$/s), while aerobic respiration consumes it quickly; consumption therefore steepens the very gradient cells climb, and aerotactic fronts form just as chemotactic fronts do~\cite{adler1966}.\keyterm{Aerotaxis}{Chemotaxis toward or away from oxygen; because respiration consumes oxygen, cells both follow and steepen the gradient they sense.} Where respiration outruns resupply, it drives oxygen to zero over micron-to-millimeter distances, creating anoxic microsites that shape dense collectives; cells stop swimming in this anoxic region, surrounded by an aerobic zone of motile cells, with aerotaxis sustaining this structure~\cite{Hokmabad2025}. Partial water saturation adds a further complication: because oxygen diffuses ${\sim}10^4\times$ faster in air than in water, the air-filled pore network becomes a rapid supply route and the gradients cells navigate are confined to the water films coating grains.

Soil presents these chemical stimuli not one at a time but in shifting combinations, and each cell must integrate the signals from its several chemoreceptors into one tumbling decision; in shallow gradients its chemotactic bias approximately reflects the sum of the per-effector biases, but competing stimuli can produce outcomes no single gradient predicts. For example, when a strong attractant that is a poor nutrient opposes a weak attractant that is a rich nutrient, \textit{E.~coli} first accumulate at the strong attractant and then, once consumption builds an opposing self-generated gradient past a critical cell density, form an ``escape band'' that migrates to the richer source, set by competition between the activity of different chemoreceptors~\cite{zhangEscapeBand2019}. Abundances shift in time as well, so whether a collective exploits a transient attractant source of size $L$---e.g., from a lysed cell or a fresh fragment of organic matter---is set by the ratio $\tau_\mathrm{chemo}/\tau_\mathrm{source}=D_a/\chi$, which compares the time to chemotax to the attractant, $\tau_\mathrm{chemo}\sim L^2/\chi$, to the time it takes for it to diffuse away, $\tau_\mathrm{source}\sim L^2/D_a$. With a small-solute diffusivity $D_a\sim10^3~\upmu$m$^2$/s, a $L\approx100~\upmu$m source lasts only $\sim10$~s, so only collectives already within $\sim10-100~\upmu$m of it can respond before the source fades away. This response time-versus-chemoeffector lifetime competition lets marine bacteria exploit ephemeral nutrient pulses~\cite{stocker2008, smriga2016}. Hindered diffusion and intermittent flow lengthen $\tau_\mathrm{source}$ in natural soil, so the estimate presented above is a conservative bound. We have, moreover, assumed quiescent pore fluid: experiments in transparent mimics show that when water flows through the pore space, heterogeneous pore velocities create local gradients that cells exploit, potentially amplifying their chemotactic response~\cite{deAnna2020}. Shear also change swimming orientation and sweeps cells toward high-shear walls, which can instead degrade chemotactic efficiency even when the gradient is unchanged~\cite{Rusconi2014}. How these effects manifest in the structured, intermittently wetted pore space of soil across textures and flows, and how they vary as that flow waxes and wanes with wetting, remains unknown.

\subsection{Collective growth is shaped by chemical gradients and flow}
A growing bacterial collective shapes, and is shaped by, the same coupling between chemical consumption and transport that directed its migration. In quiescent conditions, cellular metabolism generates nutrient gradients within the growing collective: balancing diffusion of a growth-limiting substrate against its consumption by the cells yields the same length scale $\ell_a\sim\sqrt{D_a a_\infty/(\rho\kappa)}$ identified in \S3.1, here applied to the resource the collective lives on rather than the chemical gradient it moves along. Once a collective grows past $\ell_a$, its interior starves while only an active surface layer of thickness $\sim\ell_a$ keeps growing~\cite{stewart2003,wessel2014oxygen,martinezcalvo2022,bravoVerticalGrowthDynamics2023}. Despite its rapid diffusion, oxygen has the shortest penetration depth in dense aerobic collectives, and therefore limits their growth: it is scarce in water ($a_\infty\sim0.25$~mM) and consumed quickly, so only penetrates $\ell_a\sim10^1-10^2~\upmu$m---as confirmed in natural and model soil aggregates~\cite{sexstone1985, wilpiszeski2019}. More abundant, slowly-consumed carbon sources penetrate $\ell_a\sim10^2-10^3~\upmu$m; in slow-respiring soil aggregates, and in low-density collectives in which most cells are dormant, this penetration depth can grow to be as large as tens of millimeters. The functional consequences of this metabolic stratification are still unclear, although recent work suggests that it could influence how bacterial collectives resist antibiotics, which often only target metabolically-active cells~\cite{hancock2026}, or could pattern biofilms as they develop~\cite{squyres2025single}.

This self-generated gradient does more than internally stratify bacterial collectives---it reshapes them. Whether that reshaping roughens or smooths the collective's outer boundary depends on where the active, nutrient-exposed layer of cells sits relative to that boundary. When the growing cells are themselves the nutrient-exposed surface, shape and growth feed back directly: random fluctuations at the surface of the collective enable some cells in the active surface layer to access nutrient more readily than others, generating spatial variations in surface growth that feed back on the collective's shape, ultimately giving rise to a morphological instability that causes the collective to grow in fractal, ``broccoli''-like shapes \textbf{[Figure~\ref{fig:meso}D]}~\cite{martinezcalvo2022}. The instability is the 3D counterpart to a long line of growth-front morphogenesis studied in two dimensions (2D)~\cite{benjacob1994, Wakita1994, hallatschek2007}. In the inverse case where metabolic activity is confined to a buried base of cells far from the free boundary of the collective (e.g., when a nutrient source is embedded deep inside), however, the intervening cells instead damp the surface fluctuations and the topography of the outer boundary is frozen in place as the colony grows~\cite{bravoYunker2024}. The instability and fractal-like morphogenesis was observed for densely-packed bacterial colonies in nutrient-rich conditions; in soil, by contrast, collectives can be more sparse and slowly-respiring, lengthening $\ell_a$. Moreover, it was observed in a soft hydrogel packing that yields almost freely, whereas soil is rigid and more confining---potentially resisting the differential surface growth that drives the instability, so the quenched disorder presented by soil's pore space could reshape where roughening can proceed and blunt or pin it altogether. The pressure generated by growth in such rigid confinement can be as large as $\sim10^4-10^6$~Pa~\cite{farrellMechanicallyDriven2013,contBiofilmsDeformSoft2020,delarueSelfDrivenJamming2016}, potentially feeding back on the collective itself to slow growth, trigger biofilm EPS production, raise antibiotic tolerance, and even deform/fracture the surroundings~\cite{chuSelfinducedMechanicalStress2018, monnappaGrowthConfinementPromotes2026a,delarueSelfDrivenJamming2016}. Incorporating the influence of such mechanical interactions and their feedback on cellular behavior into the theoretical framework of Ref.~\cite{martinezcalvo2022} will be a useful direction for future work.

All of the above assumes quiescent pore fluid. When water flows through the pore space, it further reshapes the growth of sessile collectives, particularly EPS-producing biofilms, via hydrodynamic stresses. In particular, at a pore throat, a biofilm faces a competition between its mechanical strength, which we propose can be approximately characterized by its yield stress $\sigma_{\text{colony}}$, and the shear stress imposed by the flow, $\sigma_{\text{flow}}$---captured by the Bingham number $\Bi_s\equiv\sigma_{\text{colony}}/\sigma_{\text{flow}}$. When $\Bi_s>1$ the cellular aggregate behaves elastically and grows as a rounded mass, while when $\Bi_s\lesssim1$ the shear exceeds the yield stress and fluidizes the biomass, causing it to be drawn downstream into a filamentous ``streamer''~\cite{kurz2022competition,Lee2023}. Once formed, the streamer bridges obstacles \textbf{[Figure~\ref{fig:meso}E]} and grows by capturing cells and biomass from the flow, clogging the network on a time scale set by that capture rather than the doubling time of the cells within it~\cite{drescher2013, Scheidweiler2024}. As noted in \S\ref{sec:pore}.2, flow in soil typically has $\dot\gamma\sim0.1-1~\mathrm{s}^{-1}$, corresponding to $\sigma_{\text{flow}}\approx\mu\dot\gamma\sim10^{-4}-10^{-3}~$Pa for pore water of viscosity $\mu\sim1~\mathrm{mPa}\cdot\mathrm{s}$, although polymers in the pore fluid can increase this range. Biofilm yield stresses, by contrast, are far larger, with $\sigma_{\text{colony}}\sim10^0-10^3~$Pa~\cite{pavlovsky2013,charltonViscoelasticityBiofilms2019,ohmuraMicrorheologyBiofilms2024}; therefore, it may be that streamer formation is rare in most of the soil pore space, and is instead confined to the coarsest pores and to strong, transient flow events, such as rainfall infiltration, irrigation, and rapid drainage, for which $\sigma_{\text{flow}}$ can reach as large as $\sim10^1-10^2$~kPa. It could also be that aggregates reach the threshold for streamer formation through their own growth as they narrow pores and thereby increase the shear stresses resulting from continued fluid flow. Indeed, in microfluidic porous media the same competition between growth-driven narrowing and flow-driven shear sets where and how fast the pore space clogs, and drives intermittency in the preferential flow paths as pores alternately seal by growth and reopen by shear~\cite{kurz2023morphogenesis, kurz2022competition}. Macroscopic signatures have been measured in soil: microbial growth lowers bulk hydraulic conductivity and reroutes flow into preferential paths~\cite{thullner2002, volkBiofilmEffectSoil2016}. What is not resolved is whether streamers, specifically, mediate this behavior inside natural soil.

\section{THE LANDSCAPE SCALE}
\label{sec:landscape}
\begin{figure*}[t]
\centering
\includegraphics[width=\textwidth]{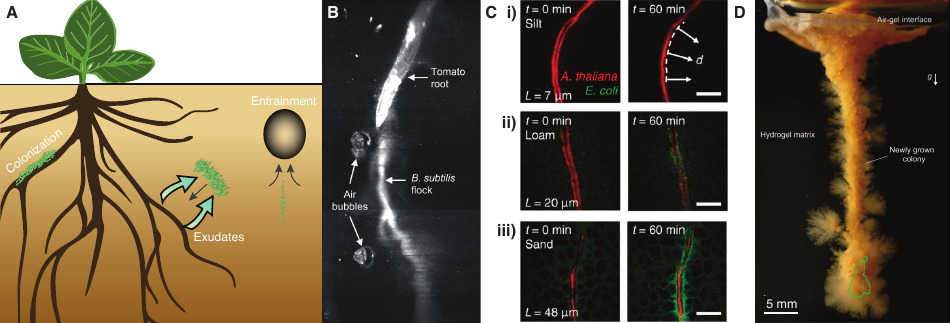}
\caption{\textbf{At the landscape scale, bacterial collectives navigate
root-imposed gradients and disperse through the soil matrix they rebuild.}
(\textbf{A}) Schematic of a landscape: a plant root secretes exudate, bacteria migrate
as a chemotactic front toward the root and colonize its surface, and a
rising biogenic bubble entrains cells and carries them vertically.
(\textbf{B}) Experimental micrograph showing \textit{B. subtilis} migrating toward a tomato
root~\cite{engelhardtNovelFormCollective2022}. [Image scale not provided.]
(\textbf{C}) Experimental micrographs showing \textit{E. coli} (green) migrating toward an
\textit{Arabidopsis thaliana} root (red) in transparent packings of three
different textures, showing that chemotactic recruitment is suppressed in finer (silt) soil textures~\cite{alharraqSoilTextureRegulates2026}. [Scale bar,~$100~\upmu$m.] (\textbf{D}) A vertical column of yeast cells formed by bubble-mediated
entrainment as a bubble rises through a transparent yield-stress granular hydrogel
matrix~\cite{hokmabad2025bubbles}. Panels (B-D) adapted with permission
from their sources.}
\label{fig:landscape}
\end{figure*}

Across a landscape ($\gtrsim10^3~\upmu$m) \textbf{[Figure~\ref{fig:landscape}A]}, external sources introduce chemical gradients over millimeters to centimeters and hours to weeks, far exceeding the scales of a single cell or a migrating front. In turn, the collective's own modifications to the solid soil matrix persist after the cells disperse or die, introducing memory to the landscape.
A collective therefore navigates gradients it did not create, reshaping them only near sources such as plant roots (\S\ref{sec:landscape}.1), and durably rebuilds the soil matrix, cementing and clogging the pore space or opening it with gas (\S\ref{sec:landscape}.2). As in the previous sections, these landscape-scale behaviors can be understood through a recurring competition between a length, time, or stress scale of the collective and one of the environment---now with the environment extending over scales far larger than the collective and carrying a history of past feedback.

\subsection{Collectives shape and respond to gradients differently across scales}
In \S\ref{sec:meso}.1, a bacterial collective consuming a chemoeffector generated a gradient over the depletion length $\ell_a$ and migrated along it as a coherent front, with the chemoeffector source treated as a single patch of approximately fixed strength. Across a landscape, however, the sources themselves---root exudates, lysed cells, decomposing litter---impose larger-scale structure of their own. The dominant source is plant roots, which continuously release a few percent of their fixed carbon as exudate over a length $L\sim1-15~\mathrm{mm}$~\cite{jonesCarbonFlowRhizosphere2009, dennis2010, kuzyakovMicrobialHotspotsHot2015a}. This process varies over a ${\sim}24$~h cycle; as a result, the population density of fast copiotrophs (doubling times $\lesssim5$~h) rises and falls near a root in phase, whereas slow bulk-soil populations (doubling times as long as $\sim$~months) integrate over many exudate cycles and respond to the time-averaged profile~\cite{tixierRootExudationCarbon2023, lopezGrowthRateDominant2023, caroHydrogenStableIsotope2023}. The exudate itself is a complex mixture of different compounds; organic acids such as malate drive strong chemotaxis and selectively recruit \textit{Bacillus subtilis}, whereas sugars are weak cues~\cite{deweertFlagellaDrivenChemotaxisExudate2002, rudrappaRootSecretedMalicAcid2008}. Exudation therefore enables plants to assemble a rhizosphere\keyterm{Rhizosphere}{The zone of soil within a few millimeters of a root, whose chemistry and microbial composition the plant actively shapes.}
community distinct from the surrounding bulk soil~\cite{bais2006, philippot2013}.

Whether cells can \emph{sense} the imposed gradient depends on the relative steepness ${\sim}1/L$, which indeed falls within the dynamic range of bacterial sensing, as described in \S\ref{sec:meso}.1. But whether the cells can appreciably \emph{reshape} that gradient depends on the competition between consumption and diffusive resupply i.e., the comparison between $\ell_a$ and $L$. We therefore ask how the self-generated-gradient picture of \S\ref{sec:meso}.1 is modified across a landscape: Does a collective still consume the chemoeffector to generate and climb a sharp gradient of its own, or does it instead respond to the gradient the plant root imposes?

Field measurements, corroborated by reaction-diffusion models~\cite{newmanWatson1977, darrah1991, raynaud2010}, reveal that the comparison between $\ell_a$ and $L$ defines three zones at increasing distance from the root: the rhizoplane (the root surface),\keyterm{Rhizoplane}{The root surface itself, where the outermost root cells meet the soil.}the near rhizosphere ($\lesssim 1~$mm away), and bulk soil ($\gtrsim 1~$mm away). At the rhizoplane, bacterial collectives form clusters with $\rho$ reaching $\sim10^{11}$~cells/mL and $\kappa\sim10^{-19}$--$10^{-18}$~mol\,cell$^{-1}$\,s$^{-1}$~\cite{vinolas2001}, giving $\ell_a\sim10^1-10^2~\upmu$m, far below $L$: consumption dominates and the collective sculpts a strong local gradient. However, these clusters are sparse, with only ${\sim}4$--$40\%$ of the root surface colonized by bacteria~\cite{danhornFuqua2007}, and exudate leaks freely past the uncolonized majority of the surface. In the near rhizosphere, cell density is enriched two to three orders of magnitude over bulk soil~\cite{kuzyakovRazavi2019}, giving $\ell_a\sim10^2-10^3~\upmu$m, comparable to the lower end of $L$: the collective both perturbs the chemical profile and responds to it. In bulk soil most cells are dormant~\cite{blagodatskayaKuzyakov2013} and $\ell_a\gtrsim10^3~\upmu$m: the collective responds to, but does not appreciably reshape, the imposed chemical gradient. Taken together, these estimates imply that a bacterial collective actively sharpens the exudate profile within ${\sim}1$~mm of a root, driving strong directed migration towards it, as confirmed by imaging in transparent soil mimics \textbf{[Figure~\ref{fig:landscape}B]}~\cite{engelhardtNovelFormCollective2022,alharraqSoilTextureRegulates2026}. This migration is texture-dependent, being suppressed by pore-scale confinement, not chemical reach, in fine-textured soil \textbf{[Figure~\ref{fig:landscape}C]}~\cite{alharraqSoilTextureRegulates2026}. By contrast, in the bulk soil beyond $\sim1$~mm, the bacteria merely respond to the shallow exudate gradient the root imposes, drifting toward roots and scattered nutrient sources only weakly.

\subsection{Collectives durably rebuild the landscape they inhabit}
Beyond navigating the imposed chemical fields of the landscape, bacterial collectives also physically rebuild the solid matrix of soil itself. For example, through their metabolism, some bacteria raise the local pH and carbonate (CaCO$_3$) concentration of the pore fluid; when the rate at which carbonate is produced is faster than the rate at which pore flow removes it, it precipitates, welding neighboring grains together for as long as years~\cite{dejong2010biomediated, dejong2013biogeochemical, alqabany2012, meng2021field}. This process can have dramatic consequences for bulk soil: converting a few percent of the pore volume to CaCO$_3$ can raise the unconfined compressive strength of a sand from $\lesssim10$~kPa to $\sim1$~MPa and lower its permeability by up to two orders of magnitude~\cite{dejong2006, dejong2013biogeochemical}. Bacterial collectives can also bind grains with secreted EPS rather than mineral, forming softer, more reversible, but far more widespread bonds that similarly raise soil's cohesion and mechanical stability while occluding pore throats and adsorbing to grain surfaces, lowering the saturated hydraulic conductivity by one to two orders of magnitude~\cite{tisdall1982, costa2018, volkBiofilmEffectSoil2016}. Beyond binding grains, a collective also seals the pore space by growing into it: the growth-versus-shear clogging of \S\ref{sec:meso}.2, compounded across many pores, reroutes water into a shrinking set of preferential channels that persist after growth slows~\cite{thullner2002}. Secreted EPS additionally governs how the soil holds and loses water, which we discuss further in \S\ref{sec:conclusion}.

Conversely, bacterial collectives can \emph{open} the solid matrix of soil through their metabolic activity. In waterlogged soil, fermentation, denitrification, and methanogenesis supersaturate the pore fluid with CO$_2$, N$_2$, and CH$_4$. When the gas production rate exceeds the rate at which dissolution, diffusion, and pore flow carry the gas away, bubbles nucleate and grow~\cite{boudreau2012}. A growing bubble can then advance through the pore space in one of two ways: it can invade the surrounding pores without moving grains, which requires overcoming the capillary entry pressure $\sigma_c\sim\gamma/\ell_{\text{pore}}$, where $\gamma\sim10^1-10^2~\mathrm{mN}~\mathrm{m}^{-1}$ is the gas--water surface tension, or it can push the grains apart, which requires overcoming the mechanical resistance of the matrix, set in soil by both its cohesive yield stress $\sigma_y$ and frictional confining stress~\cite{holtzman2010crossover, holtzman2012capillary}. When $\sigma_y/\sigma_c\gtrsim1$, such as in stiff, cohesive matrices or coarse-textured soil, the bubble invades the surrounding pores, leaving the granular skeleton intact; when $\sigma_y/\sigma_c<1$, such as in soft matrices or fine-textured soil, the bubble instead forces the surrounding matrix apart. In coarse soil, for example, $\sigma_c\lesssim0.1$~kPa, well below the cohesive strength of an EPS-bound or cemented matrix with $\sigma_y\sim1$--$10$~kPa, so gas fingers through the pore space; conversely, in a fine-textured soil, $\sigma_c\sim10^1-10^2$~kPa, exceeding the yield stress of soft sediments, $\sigma_y\sim10^0-10^2$~Pa, so the bubble displaces grains instead. In this latter case, the bubble continues to grow until its buoyant stress $\sim\Delta\rho\,g\,h$ exceeds $\approx5\sigma_y$, at which point it rises through the soil and entrains surrounding cells in its wake, sculpting long, vertical conduits that enable cells to disperse vertically over large distances \textbf{[Figure~\ref{fig:landscape}D]}~\cite{hokmabad2025bubbles,rebataLanda2012, boudreau2012, scandella2011}; here, $\Delta\rho\sim1~\mathrm{g}~\mathrm{cm}^{-3}$ is the mass density difference, $g$ is gravitational acceleration, and $h\sim10~\upmu\mathrm{m}-1~\mathrm{mm}$ is the bubble height.

Because mineral/EPS-bound grains, clogged channels, and gas conduits all persist after the cells that made them disperse or die, the soil carries a record of past biological activity~\cite{rahmati2023soil}. Capturing such landscape memory in theory will require treating the environmental state not as a fixed input but as a system whose governing parameters themselves evolve with a bacterial collective's activity.

\section{CONCLUSION}
\label{sec:conclusion}
The feedback between bacterial collectives and the soil they live in gives rise to rich, often surprising physics---their surroundings shape how the cells move, grow, and sense, and the cells reshape their habitat in turn. The phenomena we have surveyed exemplify that feedback across scales: cells hop and trap through the pore space and migrate collectively up self-generated gradients toward plant roots; grow into serpentine cables, fractal colonies, and nematically ordered, mechanically stressed aggregates under confinement, or are drawn by flow into streamers; switch lifestyles more readily as confinement concentrates quorum-sensing signals; and reshape the ground itself, cementing grains through mineral precipitation or secreted polymer and dispersing on rising biogenic gas bubbles. These examples are not exhaustive, but they underscore that life in soil demands the attention of biological physicists---both because it poses beautiful fundamental problems in its own right and because microbial life in soil underpins the fertility of our land, the fate of its carbon, and, with it, the sustainability of this planet.

As we have noted throughout, the mechanisms underlying these phenomena can be understood by comparing a bacterial length, time, stress, or energy scale set by the cells against the corresponding scale set by the soil, with the behavior of the system changing as the two become comparable. This framing could help to collapse results obtained across different model systems, predict which behavior a given soil should show from its own measured properties, and point to the measurements that are still missing for life in soil itself. However, this picture is deliberately simplified. It is clearest when one bacterial process competes with one soil process; when several act together, new behavior can arise that no single comparison captures, such as bacterial ``escape bands'' that form when exposed to opposed attractant and nutrient gradients~\cite{zhangEscapeBand2019}. We have also held the soil fully water-saturated, whereas real soil spends much of its life partly dry, its pore water retreating into thin films and disconnected pockets that reshape where cells can swim, feed, and signal~\cite{dechesne2010, orSmets2007, schulz-bohmCallingDistanceAttraction2018}. For simplicity, our scope was narrow in the biology being examined; for example, our discussion of bacterial motility focused on flagellar swimming, neglecting other motility mechanisms (e.g., twitching, gliding, and swarming). More broadly, we only discussed the dynamics of clonal bacterial collectives, neglecting interactions between different bacterial cell types, as well as the dynamics of the other important microorganisms---fungi, archaea, viruses, and protists---that share the pore space and interact through it~\cite{muir2026, martinezcalvo2023, dalcoSpatialSelforganizationMetabolism2023}. Each of these is not a gap in the framework presented here so much as an invitation to extend it.

\section*{Disclosure Statement}
The authors are not aware of any affiliations, memberships, funding, or financial holdings that might be perceived as affecting the objectivity of this review.

\begin{acknowledgments}
We thank Gözde Demirer, Ahmed Al Harraq, Doug Jerolmack, Dianne Newman, Trevor Nolan, Trent Northen, Dani Or, Victoria Orphan, Tom Shimizu, and Ned Wingreen for stimulating discussions, along with funding from the Linde Center for Global Environmental Science and Resnick Sustainability Institute at Caltech to support a soil-themed workshop, that shaped many of the ideas in this work. We also acknowledge support from National Science Foundation (NSF) grants CBET-1941716, DMR-2011750, and EF-2124863 as well as the Camille Dreyfus Teacher-Scholar and Pew Biomedical
Scholars Programs. This research was also supported in part by grant NSF PHY-2309135 to the Kavli Institute for Theoretical Physics (KITP). We used Claude (Anthropic, Opus 4.8) to assist with literature searching and reference identification, formatting of LaTeX and BibTeX entries, and copyediting and revision of the text. The authors reviewed and verified all content and take full responsibility for the final manuscript.
\end{acknowledgments}

\end{document}